# Worst Smells and Their Worst Reasons

Davide Falessi
*DICII*
*University of Rome Tor Vergata*
Rome, Italy
falessi@ing.uniroma2.it

Rick Kazman
*ITM Department*
*University of Hawaii*
Honolulu, USA
kazman@hawaii.edu

*Abstract*—Code bad smells are symptoms of poor design and implementation. There are several well-known smell types, such as large classes (aka God classes), code clones, etc. and they have been shown to lead to technical debt and hence to decrease code maintainability. Quality gates are a recent technology that prevents the automatic acceptance of push requests of code commits that have been identified as containing certain smells. However, it is a challenging activity to decide which smells should be included in the quality gate, as developers may choose to optimize short term benefits like time to market over long term benefits like maintainability. But some smells appear to provide *no* benefit to developers whatsoever and hence such smells should always be avoided. The aims of this paper are: 1) to identify "worst smells", i.e., bad smells that never have a good reason to exist, 2) to determine the frequency, change-proneness, and severity associated with worst smells, and 3) to identify the "worst reasons", i.e., the reasons for introducing these worst smells in the first place. To achieve these aims we ran a survey with 71 developers. We learned that 80 out of 314 catalogued code smells are "worst"; that is, developers agreed that these 80 smells should never exist in any code base. We then checked the frequency and change-proneness of these worst smells on 27 large Apache open-source projects. Our results show insignificant differences, in both frequency and change proneness, between worst and non-worst smells. That is to say, these smells are just as damaging as other smells, but there is never any justifiable reason to introduce them. Finally, in follow-up phone interviews with five developers we confirmed that these smells are indeed worst, and the interviewees proposed seven reasons for why they may be introduced in the first place. By explicitly identifying these seven reasons, project stakeholders can, through quality gates or reviews, ensure that such smells are never accepted in a code base, thus improving quality without compromising other goals such as agility or time to market.

## I. Introduction

It is well known that the majority of the development effort over software project's lifetime [1], and up to 80% of the total cost [2] is spent in maintenance and evolution. Therefore, it is important for most projects to develop a code base that is easy to maintain.

Code bad smells are symptoms of poor design and implementation thought and effort [3]. There are several well-known smell types, such as large classes (also known as God classes), code clones, etc. and they have been shown to lead to technical debt and hence to significantly decrease code maintainability [4], [5]. To avoid code smells, prior research in software engineering has focused on providing rules that can be automatically checked by static analysis tools. These tools can thus help to enforce quality rules such as high comment density, low code complexity, and hundreds of other best practices [6], [7]. A code smell is, in this light, a violation of a specific quality rule.

These advances in quality rules and their detection via static analyzers have greatly improved the theoretical body of knowledge—as well as the practice—of how a system should be built to facilitate software evolution. Quality gates are a relatively recent technology that prevent the automatic acceptance of push requests of code commits that have been identified as containing certain smells. Thus quality gates are a way to prevent technical debt from being inadvertently inserted into the code in the first place, which then helps a project avoid all the drama, uncertainty, risk, and cost surrounding refactoring. Refactoring is expensive and risky as any code changes have the possibility of introducing side effects and bugs [8]. Therefore it is a challenging activity to decide which smells should be included in the quality gate, as developers need to balance improvements in code quality (through refactoring) with more tangible and short-term objectives, e.g., the cost and time-to-market of customer-facing features.

Balancing these concerns is hard because improvements to a code base's maintainability are largely invisible to decision makers, and do not have an immediate impact on functionality and on the end-user. As a matter of fact, several studies have agreed that technical debt is something a project must often live with as it is important to balance code quality with other, sometimes more pressing, business goals such as time to market [7], [9]. Moreover, our previous findings from working with practitioners shows no correlation between the level of business success of a product and its level of technical debt [7]. In other words, we have observed several projects having a higher economic success than projects that demonstrably contain less technical debt. In conclusion, on one hand we know that fewer smells are "better", but on the other hand we know that in the end, this is a matter of compromises driven by practical, often extraneous concerns. Similarly we know that developers (intentionally or unintentionally) violate design and coding rules and best practices, and so technical debt inevitably accumulates over time, making the system harder to maintain [4], [5].

There are smells like "god classes" [10], also called large classes [3], that seem to organically grow in a code base without the development team ever making any explicit decision to have such classes (although there is some conflicting evidence about how harmful these classes are [4], [11]). In other words, large classes typically start small and, over the life of a project, as new features are added, they pass a threshold and they are then deemed to be "too big". In this way, we can understand that there may be good reasons why a developer would commit code that violates the "god class" rule. The developers may need to add some simple functionality to a class that is already almost too big, and they prefer to commit code that violates the size rule instead of restructuring the code, which would require significant extra effort and could even lead to new bugs. Thus, even assuming that god classes are indeed harmful, it is debatable whether strict quality gates should be used to avoid *all* god classes.

Despite these kinds of uncertainty, some smells appear to provide no benefit whatsoever and hence should *always* be avoided. Take the example of a method named "ThisIsAGreat-Method1". Is there ever any good reason not to provide a more intuitive and explanatory name for this method? Who benefits from this lazy naming? Should this code be accepted in the code base? In this paper we leverage the intuition that some smells may provide no benefit to anybody and are purely mistakes of omission or commission, and these mistakes should unequivocally be avoided. Such smells should be identified for education purposes, and strictly enforced as quality gates.

Based on this insight, the aim of this paper is threefold: 1) to identify these "worst smells", i.e., bad smells that never have a good reason to exist, 2) to investigate the frequency and change-proneness of worst smells, and 3) to identify the "worst reasons", i.e., the reasons for introducing such smells in the first place.

To investigate these questions we conducted a survey of 71 working developers. Our results show that 80 out of 314 catalogued code smells were considered as worst; i.e., the surveyed developers agreed that these 80 smells should never exist in any code base. We then checked the frequency of these worst smells in 27 large Apache open-source projects and we examined their correlation to change proneness in each project. Our results showed an insignificant difference, in both frequency and change proneness, between worst and non-worst smells. That is to say, these smells are just as harmful as other smells, but there is never any justifiable reason to introduce them. Finally, in our follow-up phone interviews with five developers we confirmed that these smells are indeed worst. The interviewees proposed seven reasons for why these smells could be introduced in the first place. By explicitly cataloguing these seven reasons, we have provided a resource so that project stakeholders can—through quality gates or reviews—ensure that such smells are never accepted in a code base. In this way a project can improve the quality of its code base without compromising other important goals such as agility or time to market.

We note that in this paper we focus on code smells as detected by SonarCloud. Some of these smells might differ from what the community generally refers to code smells (i.e., Fowler's smells [3]). Our replication package (see footnote 3) reports the specific code smells taken into account in this study.

The remainder of this paper is structured as follows. Section 2 contains information about the background and related work. Section 3 describes our study design. Section 4 details the results of our investigation. Section 5 is a discussion of results. Section 6 explains the threats to validity. And Section 7 provides conclusions and our thoughts on future directions.

## II. RELATED WORK & BACKGROUND

The term technical debt was first introduced in 1992 by Ward Cunningham [12] who was the first to coin the metaphor that represents "not-quite-right code" as a form of debt in a software system. Although Cunningham only called this "debt" initially, the term Technical Debt gradually gained broad acceptance [13], [14]; it now refers not only to impediments in the code but also in the architecture [15], [16], test [17] and social structures [18] involved in the development of software systems [19]. For instance, according to Ernst et al. [20], architectural decisions are the most important source of technical debt.

Tamburri and Di Nitto [21] analyzed an industrial case-study and found that software architecture plays a major role in causing social debt. Moreover, they provided a Goal-Question-Metric-based framework called DAHLIA to elicit some of the causes behind social debt.

Several research works have tried to characterize the specific causes of technical debt. For instance, Lim et al. [9] report an interview of 35 practitioners focused on characterizing technical debt. Their results show that most participants focused on technical debt as a result of intentional decisions to trade off competing concerns during development. This does not mean that intentional decisions are more prevalent in industry than unintentional decisions. It could be that their interview structure, the type of interviewee, or interviewee bias favored reporting of intentional debt over sloppy programming or poor developer discipline.

Brenner [22] recently reported eleven nontechnical phenomena that contribute to the formation or persistence of technical debt. However, the existence of these eleven nontechnical causes does not necessarily imply that technical causes are more prevalent in real-world projects than nontechnical causes.

Martini et al. [23] provide a taxonomy of causes of Architectural Technical Debt by conducting conducted an exploratory multiple-case embedded study in 7 sites at 5 large companies. And Xiao et al. [24] have explored the automatic detection and quantification of architectural debt, along with the negative consequences of this debt, in terms of bugs and code changes.

Tufano et al. [25] performed a substantial empirical study over the change history of 200 open source projects. Their results show that most of the smells are introduced when an artifact is created and not as a result of its evolution.

A large body of research has tackled the existence of self-admitted technical debt [26]–[29], i.e., developers explicitly reporting, in their source code comments, the existence of smells. All this research is, it should be noted, focused on the investigation of smells that are introduced intentionally and deliberately.

Very recently, Lenarduzzi et al. [30] mined 21 opensource projects to understand the level fo harmfulness of SonarQube rules and if it deffers among rules. They suggest companies carefully consider which rules they really need to apply, especially if their goal is to reduce fault-proneness. Interestingly, our view is similar but the opposite; we believe all rules shall be applied unless there is a good reason not to do so. Our paper shows there are many rules without any good reason to be violated and hence they should be applied regardless of their harmfulness.

Moreover, Baldassarre et al. [31] analyzed both diffuseness of SonarQube rules violations and accuracy of remediation time, estimated by SonarQube, to fix violations on a set of 21 open-source Java projects. Their result show that violations are highly diffused that the remediation time estimated by SonarQube is in most cases overestimated.

The primary differences between all of this prior work and ours is in the fact that we tried to link each of the identified smells, over a large data set, to a specific type of cause, to understand whether the debt was introduced with or without a good reason.

## III. EMPIRICAL STUDY DESIGN

While there is a substantial amount of research on code and architectural smells we are, as we have just described, specifically interested in characterizing those smells that are "worst" because there is never a good reason to introduce them. To better understand this topic we have devised three concrete research questions.

### A. RQ1: Which smells are worst?

In this question we aim to identify which of the many smells that have been defined in the past have no justifiable reason to exist.

*1) Target Population:* To gather opinions on worst smells we targeted our survey at a community that was clearly interested in software and software engineering: the members of the IEEE Technical Council on Software Engineering (TCSE), which is the volunteer community that serves as the voice for software engineering within the IEEE and the Computer Society, and which was chaired by the second author at the time of this research. We employed the TCSE mailing list to recruit survey participants.

*2) Smell selection:* We initially considered all 314 of the smells detected by SonarCloud for Java[1]. However, while planning the survey, it became obvious to us that the fewer smells we asked about, the higher the number of answers per smells we would receive, on average. Therefore, instead of surveying all 314 of SonarCloud's smells we first filtered out the most obvious smells; that is, we removed from the survey the smells that clearly had at least one good reason to be introduced.

To determine which smells to filter out and which to leave out we employed a pedagogical exercise as part of two sections of a third year Software Engineering course taught by the first author in Fall 2018 at the California Polytechnic State university. In this exercise approximately 60 students were divided into 12 groups. Each of these groups were asked to identify smells without reasons, through a discussion and consensus process. Each of the 314 smells had, at that point in the semester, been previously discussed in class and typically agreement was quickly reached about whether a smell had at least one good reason to be introduced or not. To identify good reasons we asked questions such as: could this smell improve the performances, security, or any other aspect of the system? As a result, we identified 133 potential worst smells; the remaining 181 had at least one good, obvious reason to be introduced.

*3) Survey:* Based on the above process, we devised a survey to elicit the opinions of practitioners about the selected worst smells. The survey was structured into four parts: Introduction, Experience, Smells Opinions, and Other. In the Introduction section we explained the research context and highlighted that "The purpose of the study is to discriminate smells that have no reason to be introduced from smells that have a good reason to be introduced. As such, there is no right or wrong answer; we just want to understand what developers (i.e., you) think."

In the Experience section we asked "How many years of experience do you have with Java Programming?" Possible answers were: Less than 1 year, 1-2 years, 2-3 Years, 3-5 years, 6-10 years, and 11+ years.

In the Smells Opinions section, for each of 11 chosen smells, drawn randomly for each respondent, from the set of 133 potential worst smells, we asked for the respondent's level of agreement with the following statement: "There is no reason for this smell to be introduced." Possible answers were, in this order: Strongly disagree, Disagree, Unsure, Agree, Strongly agree.

We put the disagreement options first (rather than the agreement options) to avoid a potential anchoring effect based on our selection of these smells as potentially worst. In addition, we did not disclose that we filtered out smells that we had previously determined to be non-worst.

We also requested that subjects "Please provide reasons (if any) why this smell could be introduced (optional)." as an optional open text response. Each smell was identified via a title, with a link to the SonarCloud explanation of the associated rule which contains a textual explanation as well as a correct and incorrect example of the rule, e.g., "A field should not duplicate the name of its containing class."[2].

*4) Data Analysis:* We performed data analysis as suggested by Kitchenham et al. [32]. We defined a smell as "worst" if

---
[1] https://sonarcloud.io/

[2] https://rules.sonarsource.com/java/RSPEC-1700

there was at least one "agree" or "strongly agree" response and no "disagree" or "strongly disagree" responses.

### B. RQ2: Are worst smells different than other smells?

Once we had determined which smells are worst we wanted to know if they somehow behave similar to normal smells. That is, we were interested to know if these worst smells have the same frequency and negative impacts on maintainability as non-worst smells.

*1) RQ2.1: Are worst smells as frequent as non-worst smells?:* The identification of worst smells and their reasons has an impact if those smells are in fact introduced in common practice. Therefore we identified 27 open-source Apache projects that are Java and Maven based. We then counted, for all classes of each release of each project, the **occurrences** of worst and non-worst smells. Note that if a smell was introduced multiple times, in one or more classes of the same release of a project, then we counted it multiple times.

To count the smells we ran SonarCloud on a local server for each of the 667 releases of the 27 open-source projects.

Our conjecture was that the raw frequency of worst smells would be higher than that of non-worst smells. This assumption was simply due to the fact that there were many more (about three times as many) non-worst smells than worst smells identified. Therefore, as described above, we also computed the **weighted occurrences** of worst and non-worst smells as the total occurrences divided by the size of the smell group. For instance, the weighted occurrences of worst smells have been computed by first computing the total violations of all worst smells in a given release of a project and then by dividing this total occurrences by the number of worst smells. In the end, as we shall describe, we determined that 80 of the 314 smells were indeed worst, and so this number was used to normalize the reported counts of worst smells. Similarly, the total occurrences of non-worst smells were normalized by dividing them by 234 (i.e., 314 - 80). These frequencies give us a proxy of the likelihood of a smell of a group to exist, regardless of the size of the group. Our resulting conjecture, given this normalization, was that worst smells would be similar to non-worst smells in terms of their weighted occurrences.

Thus, in this research question we established two null hypotheses:
- **H1₀:** The occurrences of worst smells are the same as non-worst smells.
- **H2₀:** The weighted occurrences of worst smells are the same as the weighted occurrences of non-worst smells.

Given the type of our data, the hypotheses above were tested using the Kruskal-Wallis test [33] which is similar to the more famous Anova test but does not have any assumptions about the distribution of the data. Because the non-parametric Kruskal-Wallis test is less powerful than the Anova test, it is more prone to not rejecting hypotheses when they actually need to be rejected, but on the other hand, when rejecting a hypothesis, it is more reliable than Anova. Moreover, the Kruskal-Wallis test is particularly recommended when the compared distributions are not independent. In our case, the distributions are computed over the same 667 releases of the 27 projects and hence are not independent.

In this study we use a confidence level, i.e., alpha, of 5% as is standard in most software engineering studies [34].

*2) RQ2.2: Are worst smells as change-prone as non-worst smells?:* The identification of worst smells and their reasons is important if those smells have a negative impact on a project's maintainability. Among the possible types of negative impact we chose to measure change proneness. We measured this as the number of lines touched by revisions in the project's revision history.

Therefore, for each of the 667 releases we measured the total **changed lines** across all the classes of that release. We then measured the **proportion of worst smells** versus non-worst smells in that release; that is we divided the occurrences of worst smells by the total occurrences of smells, among all code files in each release. The reasoning is that if worst smells are less harmful than non-worst smells then we would see a negative correlation between the number of changed lines and the proportion of worst smells. Thus, our conjecture is that there is no correlation between the number of changed lines and the proportion of worst smells.

One could argue that a null correlation between number of changed lines and the proportion of worst smells could be due to factors other than smells. Therefore, we also measured the correlation between number of changed lines and occurrences of smells. Our expectation is that the occurrences of smells is positively correlated with number of changed lines. If true, this would mean that smells impact change-proneness, but their type (worst or not) does not.

Therefore, in this research question we proposed two null hypotheses:
- **H3₀:** There is no correlation between the proportion of worst smells and number of changed lines of code.
- **H4₀:** There is no correlation between the occurrences of smells and number of changed lines of code.

We use the Spearman Rho as a correlation measure since it is non-parametric and hence does not have assumptions on the normality of the distributions [35].

*3) RQ2.3: Are worst smells as severe as non-worst smells?:* In this research question we measured the level of severity of a smell as reported by SonarCloud in an ordinal scale: Info, Minor, Major, Critical, and Blocker. Our assumption was that worst smells would be similar to non-worst smells in terms of their levels of severity. Therefore our null hypothesis was that:
- **H5₀:** There is no difference in the severity of worst smells versus non-worst smells.

Given the type of our data, to test this hypothesis we made use of Fisher's exact test [36].

*C. RQ3: What are the worst reasons for smells?*

We invited all the subjects who participated in the original survey to engage in a follow-up semi-structured phone interview that focused on the smells identified in our investigation into RQ1. As a result of that original survey the initial population of 133 candidate worst smells was whittled down to 80, as we described above.

During these subsequent phone interviews we first explained the research context and the general aim of the study and interview. Then we randomly selected one of the 80 smells and asked the respondents to comment on why they believe it is a smell and the possible reasons why a developer might introduce it. Finally we double-checked with the interviewee that no good reason existed to introduce the smell. The interview ended at the subject's request. On average we discussed 4 smells per interview.

We transcribed the interviews and both authors independently analyzed the answers to confirm that the smells were actually worst, according to the opinion of the interviewee. As part of our analysis we distilled a set of "worst reasons" to insert a smell into a code base, other than the previously known ones such as inexperience or time pressures. Identifying those worst reasons is important if we hope to use these results to create guidelines to prevent the introduction of worst smells (and of smells in general).

## IV. RESULTS

We now describe the answers to, and evidence for, each of our three research questions.

*A. RQ1: Which smells are worst?*

A total of 71 subjects responded to the survey invitation to the TCSE mailing list, providing a total of 1947 answers to the various questions. Table I reports the levels of programming experience, in years, of these 71 survey respondents.

According to Table I it is clear that the majority of the respondents had an appropriate level of experience to make judgements on smells.

TABLE I
LEVEL OF EXPERIENCE OF THE SURVEY RESPONDENTS.

| Level of Experience | Proportion |
|---|---|
| Less than 1 year | 12% |
| 1-2 years | 8% |
| 2-3 Years | 10% |
| 3-5 years | 7% |
| 6-10 years | 17% |
| 11+ years | 47% |

Out of the 133 candidate worst smells that we had previously identified, **80 were confirmed as worst smells**. Our hurdle to consider that a smell was confirmed as worst was that at least one respondent identified it as worst, and no other respondent found any good reason to introduce this smell. Based on this evidence we conclude that **25.5%—just over a quarter—of the 314 smells identified by SonarCloud have no good reason to be introduced in code.** Given the high number of smells, instead of reporting them here we describe them in an external file included in our replication package (see footnote 3), together with their ID, definition, a link to an external description, and their type (worst or non-worst).

*B. RQ2: Are the worst smells meaningfully different than other smells?*

In this section we discuss whether worst smells are different than other smells in a way that has important consequences for developers in practice. We look at the weighted occurrences of each smell type, the change-proneness associated with them, and their severity.

*1) RQ2.1: Are worst smells as frequent as non-worst smells?:* Figure 1 shows the occurrences of the two types of smells. Similarly, Figure 2 reports the occurrences of the two types of smells per project.

According to Figure 2, we can observe that, in some projects, the average number of occurrences of non-worst smells (over all releases) was very high; in some cases more than 200 per release. This is likely a simple effect of size: a release of these projects may contain many classes, each consisting of several hundred lines of code.

Regarding the differences between worst and non-worst smells, as can be seen in the top section of Figure 1, non-worst smells are introduced more often than worst smells. Similarly, as it can be seen on the top section of Figure 2, in each of the 27 projects, the non-worst smells are more common than worst smells. However, this can be attributed to the fact that there are three times more non-worst smells than worst smells. In fact, as can be seen on the bottom section of Figure 1, the proportion of times worst smells are introduced is higher than non-worst smells.

Similarly, as can be seen on the bottom section of Figure 2, in most of the projects the weighted occurrences of worst smells is higher than non-worst smells. We observe a large variance across projects in occurrences and weighted occurrences. This large variance can be due to the fact that some projects are much smaller than others in terms of lines of code, and hence likely in terms of occurrences and weighted occurrences. Finally, according to Figure 2, the difference in occurrences is more dramatic in some projects, e.g., *axiom* and *net*, than others, e.g., *avro* and *http*.

Table II reports the p-value of the Kruskal-Wallis on the differences of occurrences and weighted occurrences between worst and non-worst smells. According to Table II we can reject H10 as the non-worst smells are significantly higher in occurrences than worst smells. However, according to Table II we cannot reject H20; that is, we did not see a significant difference in the weighted occurrences between worst and non-worst smells.

In conclusion, **the likelihood of violating a non-worst smell is statistically similar to that of a worst smell**, but worst smells are introduced less often in absolute terms due to their smaller numbers.

*2) RQ2.2: Are worst smells as change-prone as non-worst smells?:* Figure 3 reports the number of lines changed in

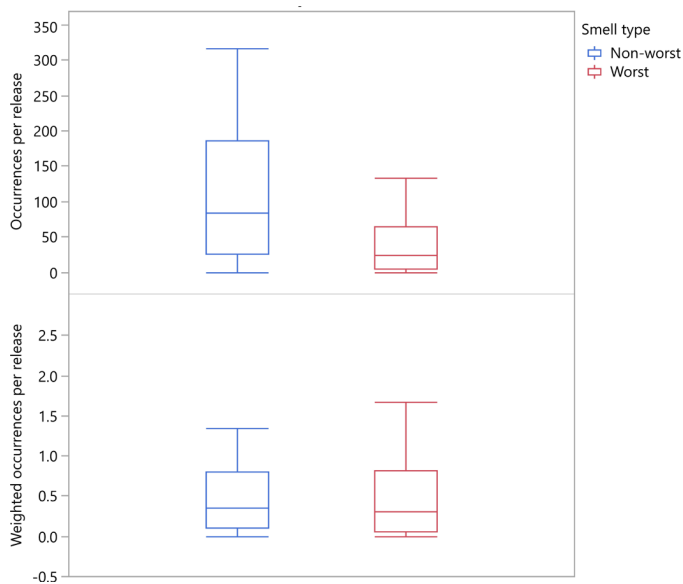

Fig. 1. Occurrences and weighted occurrences of types of smells.

TABLE II
STATISTICAL TEST RESULTS ON THE DIFFERENCE OF OCCURRENCES AND WEIGHTED OCCURRENCES BETWEEN WORST AND NON-WORST SMELLS.

| Variable | Kruskal-Wallis p-value |
| --- | --- |
| Occurrences of smells | 0.0035 |
| Weighted occurrences of smells | 0.8441 |

a release for a specific proportion of non-worst smells. Table III reports the correlation analysis between number of lines changed and percent of worst smells introduced and occurrences of smells. According to Figure 3 there is no evident trend between the number of lines changed in a release and the proportion of worst smells introduced in that release. Moreover, according to Figure 3 there is a very wide distribution of values. This is reasonable considering the several characteristics that can impact change proneness like the number of requirements implemented in, and the length of, the release. According to Table III the percent of worst smells introduced is essentially uncorrelated to changed lines in a release. As expected, according to Table III the total occurrences of smells is positively correlated with changed lines of code in a release; i.e., the higher the technical debt—as measured by SonarCloud smells—the higher the lines of code changed. Note that the correlation is statistically significant (i.e., p-value less than 0.05) in both cases thus suggesting the above levels of correlation are reliable and therefore that we can reject both H30 and H40. In conclusion, **there is no major difference between worst and non-worst smells in terms of their effects on changed lines of code in a project release**.

*3) RQ2.3: Are worst smells as severe as non-worst smells?:* Figure 4 reports the distribution of smells types over different SonarCloud severity levels. According to Figure 4 the worst smells appear to have, on average, a lower severity than non-worst smells even if the worst smells have the higher proportion of the highest severity level, i.e., blocker. Table IV reports the statistical test results on the difference of severity levels between worst and non-worst smells. According to Table IV there is indeed a significant difference on severity levels between worst and non-worst smells and therefore we can reject H50. However, we recommend care in considering one type of smell more severe than another as some severity levels can be more distant than others; in other words, in some contexts only blocker smells may be important.

TABLE III
CORRELATIONS BETWEEN NUMBER OF LINES OF CODE CHANGED, PERCENT OF WORST SMELLS INTRODUCED, AND OCCURRENCES OF SMELLS.

| Variable | Spearman's Rho | P-value |
| --- | --- | --- |
| % Worst Smells occurrences | 0.086 | 0.0463 |
| Occurrences | 0.2205 | 0.0001 |

TABLE IV
STATISTICAL TEST RESULTS FOR THE DIFFERENCE OF SEVERITY LEVELS BETWEEN WORST AND NON-WORST SMELLS.

| Variable | Fisher's exact P-value |
| --- | --- |
| % Worst versus non-worst smells | 0.0392 |

### C. RQ3: What are the worst reasons?

40 subjects out of the 71 subjects that responded to the survey provided their consent to be contacted again for a short interview. Five subjects, out of these 40, engaged in a phone interview with us. Some subjects discussed a single smell during their interview, while others discussed up to five smells. As a result, a total of 20 smells were discussed by these five subjects during the phone interviews. The interviews have been transcribed and analyzed by both authors. Our qualitative analysis shows that all 20 of these smells were indeed worst. That is, the practitioners confirmed that these smells had no good reason to ever exist in a code base.

We note that smells in general may have many reasons to come into existence. For example, a god class may be created intentionally, or it may be created unintentionally and "innocently" as the code evolves and many developers work on it, adding features and fixing bugs. However, the worst smells have the unique characteristic of never having any good reason to exist.

By analyzing the transcripts we came to a consensus on seven generic reasons for a worst smell to be created. These are briefly described in Table V.

## V. DISCUSSION

As we discussed in the introduction, there are legitimate reasons, to varying degrees, for introducing smells into a code base. An oft-cited reason is the push towards a deadline, where developers feel that they can not afford the luxury of "doing the right thing", and so they take shortcuts [14]. In some cases that debt is consciously tracked and "repaid" after the release, but frequently this is not the case and debt simply accumulates over the lifetime of the project.

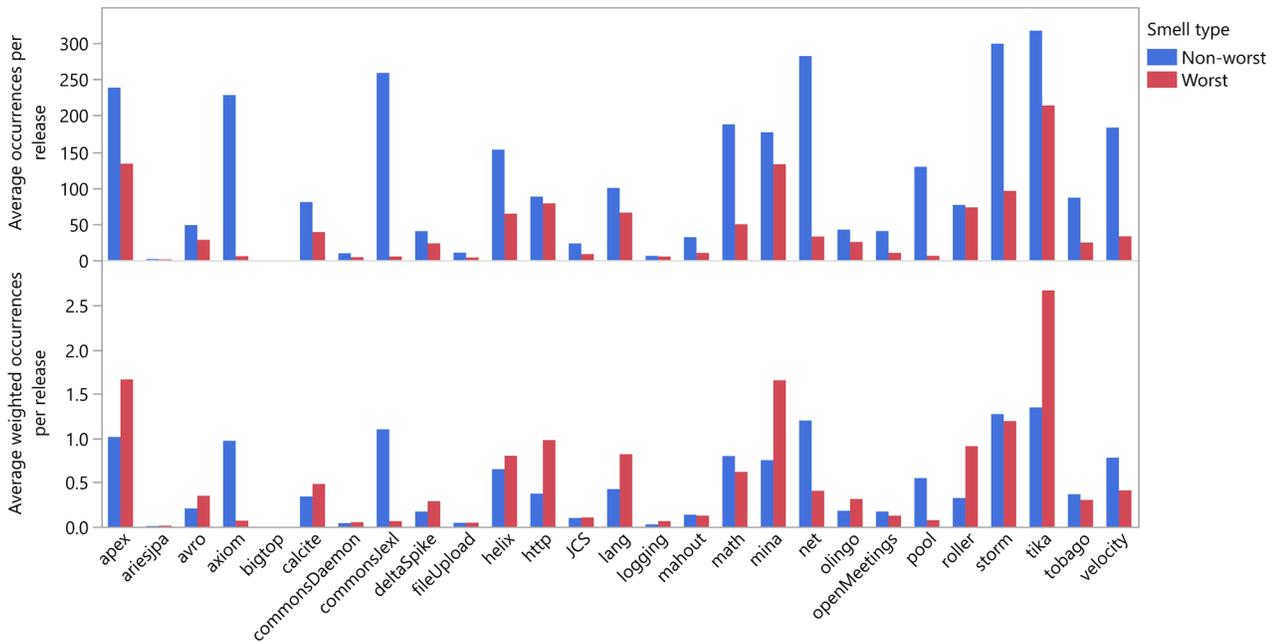

Fig. 2. Occurrences and weighted occurrences of types of smells per project.

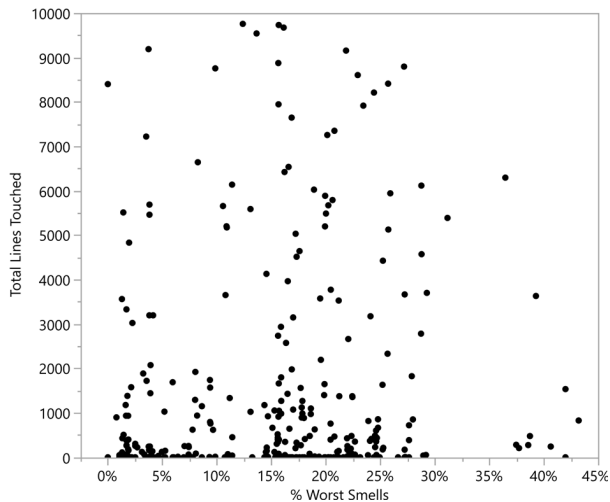

Fig. 3. Number of lines of code changed in a release versus proportion of worst smells.

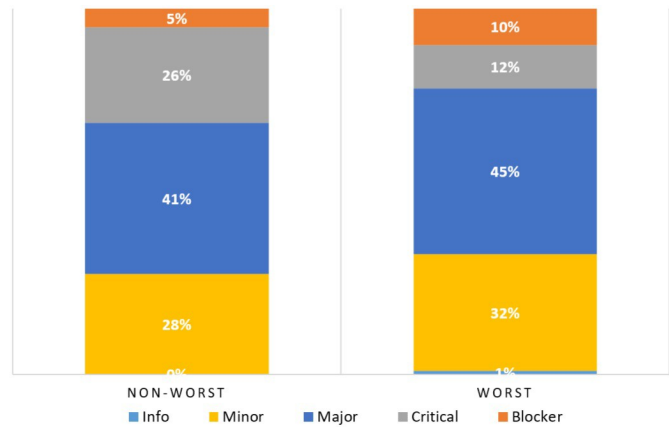

Fig. 4. Distribution of smell types over SonarCloud severity levels.

In our study, however, we wanted to distinguish the above-mentioned forms of debt which, while perhaps dangerous to project health in the long run, can be justified in the short run, from smells whose introduction can never be justified. These are, of course, what we have termed worst smells. Given that our results (specifically RQ2) demonstrate that these worst smells are not so different than non-worst smells, but their introduction is never warranted, it seems reasonable for a project to make its best efforts—via education, coding standards, quality gates, and code reviews—to ensure that they are never introduced into a code base. In practice, developers can customize their quality gates by considering our identified 80 worst smells (and rejecting commits that contain such smells).

The seven categories of reasons for these smells, presented in Table V provide a guide to a project leader in terms of how to avoid these smells in the first place. For example, the "Lack of Knowledge", "Improper language transfer assumption", and "Intentional mistake" smells can be warded off by strategies such as having clear project standards reinforced by better training of developers (with a specific focus on these smells), by project role models who lead by example, and by buddy-

| Title | Definition | ID of Good Example | Title of Good Example |
| --- | --- | --- | --- |
| Lack of knowledge | A better solution is available but it is not known by the developer who, if they were aware of it, would have little or no reason to not use it. E.g., using modifiers in the correct order. | 1124 | Modifiers should be declared in the correct order. |
| Improper language transfer assumption | The logic behind the code would make sense in another language but not in the current (Java) language. E.g., implementing some sort of garbage collector in Java may be less optimal than the standard Java garbage collector but required in C/C++. | 1113 | The Object.finalize() method should not be overriden. |
| Committing preliminary code | The code was preliminary. and should have been removed before being committed to production. E.g., commented out code such as temporary code used in development as a prototype or for testing. | 125 | Sections of code should not be commented out. |
| Neglecting a simpler solution | The code would be okay if complemented by a lot of explanation. However a simpler solution is available which is more logical and hence does not require lengthy explanation. E.g., missing an unconditional break in the final switch case. The absence of a break can be intentional (to be documented) or it can be a mistake. | 128 | "Switch cases should end with an unconditional ""break"" statement". |
| Unintentional mistake | This is an unintentional mistake made because the developer was not thinking clearly. i.e., an error in judgement rather than a lack of knowledge. E.g., using string concatenation in a loop rather than StringBuilder will degrade performance. | 1643 | Strings should not be concatenated using '+' in a loop. |
| Regression oversight | A piece of code makes no sense after a modification. E.g., an empty loop (which before was not empty). | 108 | Nested blocks of code should not be left empty. |
| Intentional mistake | The programmer was lazy and did not want to do things right and hence intentionally used a shortcut saving a few clicks. E.g., failing to use the ""@Override"" annotation to highlight that a method is being overridden does not save much time." | 1161 | Override should be used on overriding and implementing methods |

TABLE V
THE "WORST REASONS" TO HAVE WORST SMELLS.

.

pairing.

The "Regression oversight" and "Committing preliminary code" smells, while not necessarily automatically detectable, can be easily caught with code reviews, as long as the identification of such smells is clearly identified as one of the objectives of the reviews.

The final two smells—"Unintentional mistake" and "Neglecting a simpler solution" can only be addressed by a combination of the two above approaches: education and code reviews. Because these are cases where the developer was not consciously aware of their mistake, there is an obvious educational opportunity here. But until the developer's consciousness has been raised, code reviews are warranted. Moreover, code reviews can and should be augmented by analysis tools (such as the one used in this study), so that developers can get rapid feedback when they attempt to commit code with a worst smell.

Finally, we believe our results provide an impact to teaching good code habits. The entire spectrum of 314 smells represent a large set of rules to obey—far too large for any practical purpose! No human could possibly remember 314 rules. Our differentiation of smells into worst or non-worst categories, and our analysis of the reasons for worst smells, provides to students a level of understanding that is deeper than simply looking at and trying to internalize each rule in isolation. Students can better relate to, and reflect on, the impacts of a low level code snippet in terms of the required tradeoffs between long-term quality goals such as maintainability and short terms goals such as time to market.

## VI. THREATS TO VALIDITY

In this section, we discuss the threats to validity posed by the research that we have prosecuted here. Specifically we focus on construct, internal, conclusion, and external validity threats [37].

*Construct validity* threats are related to the relationship between theory and observation. It could be that our way of posing questions during the survey (RQ1) impacted the identification of worst smells. To attempt to remediate this threat we checked a randomly chosen set of the identified worst smells via our interviews (RQ3).

*Internal validity* threats concern variables internal to our study and not considered in our experiment that could influence our observations on the dependent variable. We see no threat to internal validity since the set of 314 smells were pre-existing. These smells are largely well known and they are regularly injected by practitioners. This is supported by the large number of occurrences of the smells in practice, as discussed in our analysis of RQ2.1. One main threat to internal validity is in the identification of the 80 worst smells since: 1) the subjects identifying the initial set of potential worst smells were students and not practitioners, and 2) the survey, like all surveys, is not perfect. To respond to this threat we do not claim that the set of 80 identified worst smells is complete. What we do claim is that they are correct, i.e., that the 80 smells identified in this study are indeed "worst". Moreover, regarding the reliability of students in empirical studies, we note that students cannot in general be judged as less reliable than professionals [38].

*Conclusion validity* threats surround issues that affect the ability to draw accurate conclusions about relations between the treatments and the outcome of an experiment. We note that *absence of evidence is not evidence of absence*. In our case we need to be careful when analyzing the results of our RQ2. The lack of evidence on differences between worst and non-worst smells in terms of occurrences, change proneness and severity can not be considered to be conclusive evidence of no difference in actuality.

*External validity* concerns the extent to which the research elements (subjects, artifacts, etc.) are representative of actual elements [37]. We chose, as the projects that we analyzed, just Apache projects. So this choice might have introduced a selection bias and Apache projects might in fact be qualitatively different than other projects. But at this point the Apache ecosystem is so large and diverse that the risk here seems minimal.

Finally, to encourage replicability we have provided a complete replication package[3] and we sincerely hope that others will attempt to replicate and extend our findings.

## VII. CONCLUSION

By means of a classroom exercise followed by a survey with 71 software development practitioners we identified 80 "worst smells"—smells that have no good reason to exist. We then checked possible differences between worst and non-worst smells in terms of their frequencies of occurrence and their correlations to change proneness using 667 releases of 27 large Apache open-source projects as our test dataset. Afterwards, we checked for possible differences between worst and non-worst smells in terms of their severity levels, as provided by SonarCloud.

Our results showed insignificant differences, in both frequency and change proneness, between worst and non-worst smells. That is to say, these smells are just as frequent and just as damaging as other smells, but there is never any justifiable reason to introduce them. Worst smells seem to be similar to non-worst smells also in terms of their severity: the worst smells appear to have, on average, a lower severity than non-worst smells but the worst smells have a higher proportion of the highest severity level.

Finally, our follow-up phone interviews with five developers provided confirmatory evidence that these smells are indeed worst. During the course of the interviews these interviewees collectively proposed seven distinct reasons for why these smells may be introduced in the first place. By explicitly identifying these seven reasons, project decision-makers can, through quality gates, education, and reviews, ensure that such smells are never unintentionally accepted in a code base, thus improving quality without compromising other goals such as agility or time to market.

## VIII. ACKNOWLEDGEMENT

We would like to thank the many software engineering students of the California Polytechnic State University that have been involved in this study. Kazman has been supported in this research by National Science Foundation grants CNS-1823214 and CNS-1817267.

[3]https://doi.org/10.5281/zenodo.4270178